\documentclass[twocolumn]{autart}    % Enable this line and disable the 
\usepackage{graphicx, color}          % Include this line if your 
\usepackage{amsmath, amssymb, cite}%mathptmx
\usepackage{bbm}
\usepackage{amsfonts}
\usepackage{booktabs}
\usepackage{dsfont}
\usepackage{siunitx}
\usepackage{balance}

\newtheorem{secthm}{Theorem}[section]
\newtheorem{seccor}[secthm]{Corollary}

\newtheorem{secex}[secthm]{Example}

\newtheorem{secdefn}[secthm]{Definition}
\newtheorem{secrem}[secthm]{Remark}
\newtheorem{secasm}[secthm]{Assumption}

\newcommand{\cK} { {\mathcal K}}
\newcommand{\bR} { {\mathbb R}}
\newcommand{\1}{\mbox{1}\hspace{-0.25em}\mbox{l}}
\def\red{\hfill $\lhd$}

\usepackage{tikz}
\usetikzlibrary{arrows,automata}
\usetikzlibrary{arrows,shapes,backgrounds,calc,positioning,patterns}
\usetikzlibrary{calc,arrows,shapes,backgrounds,calc,positioning,patterns,decorations.pathmorphing,decorations.markings,mindmap,trees}
\tikzstyle{block} = [draw, rectangle, minimum height=2em, minimum
width=4em] \tikzstyle{sum} = [draw, fill=blue!20, circle, node
distance=1cm] \tikzstyle{input} = [coordinate] \tikzstyle{output} =
[coordinate] \tikzstyle{pinstyle} = [pin edge={to-,thin,black}]
\usepackage[american,cute inductors,smartlabels]{circuitikz}
\usetikzlibrary{arrows,automata}
\usepackage[american,cuteinductors,smartlabels]{circuitikz}
\usetikzlibrary{calc}
\ctikzset{bipoles/thickness=1} \ctikzset{bipoles/length=0.8cm}
\ctikzset{bipoles/diode/height=.375}
\ctikzset{bipoles/diode/width=.3}
\ctikzset{tripoles/thyristor/height=.8}
\ctikzset{tripoles/thyristor/width=1}
\ctikzset{bipoles/vsourceam/height/.initial=.7}
\ctikzset{bipoles/vsourceam/width/.initial=.7} \tikzstyle{every
node}=[font=\small] \tikzstyle{every path}=[line width=0.8pt,line
cap=round,line join=round]

\begin{document}

\begin{frontmatter}

\title{Krasovskii and Shifted Passivity Based Output Consensus\thanksref{footnoteinfo}} % Title, preferably not more 

\thanks[footnoteinfo]{%This paper was not presented at any IFAC meeting. 
This work was supported in part by JSPS KAKENHI Grant Number JP21K14185. }

\author[JP]{Yu Kawano}\ead{ykawano@hiroshima-u.ac.ip},  
\author[IT,NL]{Michele Cucuzzella}\ead{michele.cucuzzella@unipv.it},
\author[CN]{Shuai Feng}\ead{s.feng@njust.edu.cn},
\author[NL]{Jacquelien M.A. Scherpen}\ead{j.m.a.scherpen@rug.nl}

\address[JP]{Graduate School of Advanced Science and Engineering, Hiroshima University, Higashi-Hiroshima 739-8527, Japan}
\address[IT]{Department of Electrical, Computer and Biomedical Engineering, University of Pavia, 27100 Pavia PV, Italy}
\address[CN]{School of Automation, Nanjing University of Science and Technology, Nanjing 210094, China}
\address[NL]{Jan C. Willems Center for Systems and Control, ENTEG, Faculty of Science and Engineering, University of Groningen, 9747 AG Groningen, the Netherlands}
          
                             % keyword list or with the 
                                          % help of the Automatica 
                                          % keyword wizard

\begin{abstract}% Abstract of not more than 200 words.
Motivated by current sharing in power networks, 
we consider a class of output consensus (also called agreement) problems for nonlinear systems, where the consensus value is determined by external disturbances, e.g., power demand.
This output consensus problem is solved by a simple distributed output feedback controller if 
a system is either Krasovskii or shifted passive, which is the only essential requirement.
The effectiveness of the proposed controller is shown in simulation on an islanded DC power network.
\end{abstract}

\end{frontmatter}

\section{Introduction}
Steering some variables to a common value is called an agreement problem.
Along with massive research attentions of network systems, 
agreement problems have been studied in various contexts such as
distributed optimization~\cite{CNS:14, HCI:18,CenedeseTAC} and synchronization~\cite{QKP:21, QKA:21} to name a few.
Our interest in this paper is output consensus (also called output agreement) under external disturbances.
This problem is motivated by current sharing for balancing demand and supply in power networks \cite{CTD:18,FCS:20},
where currents and demands are modeled as outputs and external disturbances, respectively.

%\subsection{Literature review}

Various physical systems including the aforementioned power networks possess passivity properties.
Passivity and its variant concepts have already been witnessed as useful tools for agreement; 
see, e.g., \cite{Arcak:07, SAS:10, SM:13, BZA:14, BD:15, MD:17, HCI:18}.
In particular, the works \cite{BZA:14, BD:15, MD:17} 
have studied output consensus problems based on shifted passivity.
The common problem formulation in these papers is that 
passive node dynamics are interconnected by special edge dynamics such that
the networked interconnection naturally possesses an output consensus property, and 
passivity is used as a tool for \emph{analysis}.
However, many physical systems such as DC microgrids \cite{CTD:18,FCS:20} do not have such edge dynamics.

Regarding control design, for linear DC microgrids, the paper \cite{CTD:18} provides an output consensus controller without 
explicitly utilizing passivity.
A preliminary version \cite{FKC:22} of this paper gives a shifted passivity based output consensus controller for linear port-Hamiltonian systems, but not for nonlinear systems.
In summary, a passivity based control framework for output consensus is still missing for 
general nonlinear network systems, including nonlinear DC microgrids.

\subsection{Contribution}

In this paper, we employ passivity as a tool for output consensus \emph{control} under external disturbances.
As passivity concepts, we focus on Krasovskii passivity \cite{KKS:20} (also called $\delta$-passivity \cite{SA:21}) and shifted passivity \cite{JOG:07,KKS:20},
which are different properties for nonlinear systems in general \cite{KKS:20}.
We show that a simple distributed output feedback controller solves 
the output consensus problem if
a nonlinear network system is either Kraosvskii or shifted passive, regardless of the structure of the edges dynamics.
Namely, we reveal that imposing special edge dynamics is not an essential requirement to achieve output consensus when
one takes control design into account.
Moreover, the proposed controller can handle weighted output consensus and 
partial output consensus problems also.

The main contributions of our approach can be summarized as follows: 

\begin{enumerate}
\item The main focus of \cite{BZA:14, BD:15, MD:17} is consensus analysis under special edge dynamics. In contrast, we design an output consensus controller based on Krasovskii or shifted passivity, which can handle a wider class of nonlinear network systems such as the DC microgrids in \cite{CTD:18,FCS:20}, and weighted output consensus has not been considered in \cite{BZA:14, BD:15, MD:17}.
    \item Krasovskii passivity has not been used before for output consensus control or even analysis. The aforementioned papers \cite{BZA:14, BD:15, MD:17} and a preliminary version \cite{FKC:22} for linear port-Hamiltonian systems utilize shifted passivity, but not Krasovskii passivity.
    One of the advantages of utilizing Krasovskii passivity is that we do not need to assume the existence of an equilibrium point of the closed-loop system beforehand in contrast to shifted passivity.
    \item Shifted passivity based output consensus control design for nonlinear networks is also new contribution of this paper. One of the advantages of utilizing shifted passivity is the ease of dealing with time-varying disturbances. This has been partly observed in \cite{BD:15, MD:17} for output consensus analysis, under the assumption that disturbances are generated by Sylvester-type equations for output regulation. In this paper, we do not assume this. Namely, we do not require information of disturbances for control design.
\end{enumerate}

%It is worth emphasizing that there has been no Krasovskii passivity based output consensus controller
%before this paper.

Since Krasovskii and shifted passivity are different properties, 
there is possibility to enlarge the class of systems for which the results in 
\cite{BZA:14, BD:15, MD:17} are applicable by revisiting these results from the viewpoint of 
Krasovskii passivity.
As a relevant concept of Krasovskii passivity, differential passivity \cite{FSS:13,Schaft:13} has been known.
The proposed  Krasovskii passivity based controller is also applicable to differentially passive systems
because differential passivity implies Krasovskii passivity \cite[Theorem 2.9]{KKS:20}.

\subsection{Organization}

The remainder of this paper is organized as follows.
In Section~\ref{ME:sec}, we consider current sharing for a nonlinear islanded DC power network
as a motivating example for output consensus under external disturbances.
In Section~\ref{OCL:sec}, we use linear systems to expose the main ideas of our approaches.
In Sections~\ref{KP:sec} and~\ref{SP:sec}, we show that the proposed controller solves 
the output consensus problem for Krasovskii and shifted passive systems, respectively.
Also, in Section \ref{sec:sim} the proposed controllers are applied to the DC power network in the motivating example and tested in simulation.
Finally, Section~\ref{Con:sec} concludes this paper.

{\it Notation:}
The set of real numbers is denoted by $\bR$.
The $n$-dimensional vector whose all components are $1$ is denoted by $\1_n$.
The $n \times n$ identity matrix is denoted by $I_n$.
For a full column rank real matrix $A$, its Moore–Penrose inverse is denoted by $A^+ := (A^\top A)^{-1} A^\top$.
For $P \in \bR^{n \times n}$, $P \succ 0$ ($P \succeq 0$) means that $P$ is
symmetric and positive (semi) definite.
For $x \in \bR^n$, its  Euclidean norm weighted by $P  \succ 0$ is denoted by $| x |_P := \sqrt{x^\top P x }$.
If $P = I_n$, this is simply described by $| x |$.
A continuous function $\alpha: [0, r) \to [0, \infty)$ is said to be of class $\cK$ if $\alpha (0) = 0$ and
$\alpha$ is strictly increasing. 
Moreover, this is said to be of class $\cK_\infty$ if $r = \infty$ and $\lim_{r \to \infty} \alpha (r) = \infty$.
For a scalar-valued function $V:\bR^n \to \bR$, the column vector-valued function consisting of its partial derivatives
is denoted by $\nabla V(x) := [\begin{matrix} \partial V/\partial x_1 & \cdots & \partial V/\partial x_n \end{matrix}]^\top (x)$.

\section{Motivating Example}\label{ME:sec}

\begin{figure}[t]
    	\begin{center}
    		\begin{circuitikz}[scale=0.95,transform shape]
    			\ctikzset{current/distance=1}
    			\draw
    			% transformators i and j
    			node[] (Ti) at (0,0) {}
    			node[] (Tj) at ($(5.4,0)$) {}
    			% Buck i
    			node[] (Aibattery) at ([xshift=-4.5cm,yshift=0.9cm]Ti) {}
    			node[] (Bibattery) at ([xshift=-4.5cm,yshift=-0.9cm]Ti) {}
    			node[] (Ai) at ($(Aibattery)+(0,0.2)$) {}
    			node[] (Bi) at ($(Bibattery)+(0,-0.2)$) {}
    			($(Ai)+(-0.0005,0)$) to [R, l={$R_{i}$}] ($(Ai)+(1.7,0)$) {}
    			($(Ai)+(1.7,0)$) to [short,i_={$\dfrac{\varphi_i}{L_i}$}]($(Ai)+(1.701,0)$){}
    			($(Ai)+(1.701,0)$) to [L, l={$L_{i}$}] ($(Ai)+(3,0)$){}
    			to [short, l={}]($(Ti)+(0,1.1)$){}
    			($(Bi)+(-0.0005,0)$) to [short] ($(Ti)+(0,-1.1)$);
    			\draw
    			($(Ai)$) to []($(Aibattery)+(0,0)$)to [V_=$u_i$]($(Bi)$)
    			% PCC-i
    			($(Ti)+(-1.3,1.1)$) node[anchor=south]{{$\dfrac{q_i}{C_i}$}}
    			($(Ti)+(-1.3,1.1)$) node[ocirc](PCCi){}
    			($(Ti)+(-.3,1.1)$) to [short,i>={$I_{Li}(q_i)$}]($(Ti)+(-.3,0.5)$)to [I]($(Ti)+(-.3,-1.1)$)
    			($(Ti)+(-1.3,1.1)$) to [C, l_={$C_{i}$}] ($(Ti)+(-1.3,-1.1)$)
    			% line
    			($(Ti)+(2.,1.1)$) to [short,i_={$\dfrac{\varphi_{tk}}{L_{tk}}$}] ($(Ti)+(2.2,1.1)$)
    			($(Ti)+(0,1.1)$)--($(Ti)+(.6,1.1)$) to [R, l={$R_{tk}$}] 
    			($(Ti)+(2.5,1.1)$) {} to [L, l={{$L_{tk}$}}, color=black]($(Tj)+(-2.2,1.1)$){}
    			($(Tj)+(-2.2,1.1)$) to [short]  ($(Ti)+(3.4,1.1)$)
    			($(Ti)+(0,-1.1)$) to [short] ($(Ti)+(3.4,-1.1)$);
    			\draw
    			node [rectangle,draw,minimum width=6.1cm,minimum height=3.4cm,dashed,color=gray,label=\textbf{Node $i$},densely dashed, rounded corners] (DGUi) at ($0.5*(Aibattery)+0.5*(Bibattery)+(2.25,0.2)$) {}
    			node [rectangle,draw,minimum width=2.2cm,minimum height=3.4cm,dashed,color=gray,label=\textbf{Line $k$},densely dashed, rounded corners] (DGUi) at ($0.5*(Aibattery)+0.5*(Bibattery)+(6.65,0.2)$) {};
    		\end{circuitikz}
    		\caption{Electrical scheme of node $i \in \mathcal{V}$ and transmission line $k \in \mathcal{E}$, where $I_{Li}(q_i):= G_{Li}^\star\frac{q_i}{C_i} + I_{Li}^\star + \frac{C_i}{q_i}P_{Li}^\star$.}
    		\label{fig:networks}
    	\end{center}
    \end{figure}
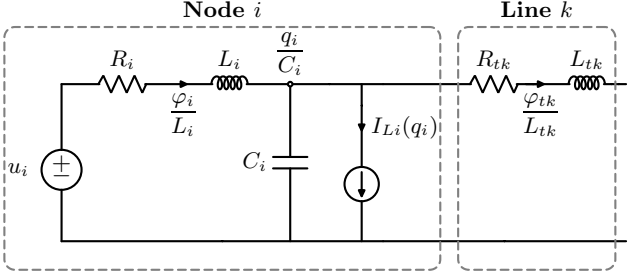

Consider an islanded DC power network model \cite{FCS:20} with $\nu$ nodes and $\mu$ edges, described by
\begin{align}\label{sys:DC}
&\dot x = f(x) + g u  + d, \\
&x := 
\begin{bmatrix}
\varphi^\top & q^\top & \varphi_t^\top
\end{bmatrix}^\top, \nonumber\\
& f(x) := (\mathcal{J} - \mathcal{R}) \nabla \mathcal{H} (x)
-
\begin{bmatrix}
0 \\ I_L^\star + {\rm diag} (C^{-1} q)^{-1} P_L^\star \\ 0
\end{bmatrix},
\nonumber\\
& \mathcal{J} :=
\begin{bmatrix}
0 & - I_n & 0 \\
I_n & 0 & D \\
0 & - D^\top & 0
\end{bmatrix}, \;
\mathcal{R} := 
\begin{bmatrix}
R & 0& 0 \\
0 & G_L^\star & 0\\
0 & 0 & R_t
\end{bmatrix}, \;
g :=\begin{bmatrix}
I_n \\ 0 \\ 0
\end{bmatrix},
\nonumber\\
& \mathcal{H} (x) := \left( | \varphi |_{L^{-1}}^2 + | q |_{C^{-1}}^2 + | \varphi_t |_{L_t^{-1}}^2 \right) /2,
\nonumber
\end{align}
where $\varphi, q \in \bR^\nu$ and $\varphi_t \in \bR^\mu$ are state variables denoting, respectively, the flux and charge of the network's nodes and the flux associated with the transmission lines interconnecting the nodes, while $u \in \bR^\nu$ denotes the control input. 
The matrices $R, R_t, L, L_t, C \succ 0$ are diagonal and have appropriate dimensions, while
$G_L^\star \in \bR^{\nu \times \nu}$, $I_L^\star, P_L^\star \in \bR^\nu$, and $d \in \bR^{2\nu + \mu}$ are unknown; see Figure \ref{fig:networks} and Table \ref{tab:symbols} for the meaning of the used symbols. The incidence matrix $D \in \bR^{\nu \times \mu}$ describes the network topology. %\textcolor{blue}{(If $\phi$ is a $n$ by 1 vector, it implies we have $n$ nodes. Similarly, we have $m$ edges. Thus, $Q , K, C\in \mathbb R ^{n \times n}$ and $Q_t, K_t \in \mathbb R ^{m \times m}$. $I_L^\star$ I guess it is the load, maybe $M I_L^\star$, where $M$ is a matrix? The term ${\rm diag} (C^{-1} q)^{-1} P_L^\star$ I cannot figure it out its representation.) }

\begin{table}[t]
\begin{center}
\caption{Description of the used symbols}\label{tab:symbols}
\begin{tabular}{clcl}
\toprule
$\varphi$						& Flux (node)&$G^\star_L$						& Load conductance\\
$q$						& Charge &$I^\star_L$						& Load current\\
$\varphi_t$ 						& Flux (power line)&$P^\star_L$						& Load power\\

$u$						& Control input &$R, L, C$						& Filter parameters\\
$d$					& Disturbance &$R_t, L_t$						& Line parameters\\
\bottomrule
\end{tabular}
\end{center}
\end{table}

To improve the generation efficiency, it is generally desired in DC microgrids that the total current demand is shared among all the nodes (current sharing) \cite{CTD:18}, i.e., for some $\alpha \in \bR$,
\begin{align}\label{ob:DC}
\lim_{t \to \infty} \varphi_i (t)/L_i = \alpha, 
\quad 
\forall i = 1,\dots,\nu.
\end{align}
Inspired by \cite{TCC:19} and based on the passivity properties of the DC network \eqref{sys:DC}, we will later reveal that \eqref{ob:DC} is achieved by 
the following simple distributed output feedback controller:
\begin{align}
\dot u &= - \mathcal{L} y, 
\label{FB:DC}\\
y &=g^\top \nabla \mathcal{H}(x) =[\begin{matrix}\varphi_1 (t)/L_1 & \cdots & \varphi_{\nu} (t)/L_{\nu}\end{matrix}]^\top,
\label{y:DC}
\end{align}
where $\mathcal{L}\succeq 0$ is such that ${\rm ker}(\mathcal{L}) = {\rm span}\{ \1_\nu\}$. For example, $\mathcal{L}$ can be the Laplacian matrix associated with a connected and undirected communication graph.

We note that the DC network \eqref{sys:DC} is not a conventional port-Hamiltonian system 
because of the term $-I_L^\star + {\rm diag} (C^{-1} q)^{-1} P_L^\star$ in $f(x)$.
However, we will show later that the system  \eqref{sys:DC} possesses two different passivity properties: 
1) Krasovskii passivity \cite{KKS:20} and 2) shifted passivity \cite{JOG:07,KKS:20}.
In general, these two properties are different for nonlinear systems \cite{KKS:20}.
In the rest of this paper, we show that if a system is either Krasovskii or shifted passive, 
then \eqref{FB:DC} is a controller achieving output consensus with respect to the passive output.
For the DC network \eqref{sys:DC}, the passive output for both Krasovskii and shifted passivity is $y$ in \eqref{y:DC}.

\begin{secrem}
For the DC network \eqref{sys:DC}, the results in \cite{BZA:14, BD:15, MD:17} for output consensus analysis are not applicable. 
These papers impose special edge dynamics such that the interconnected system naturally possesses  a consensus property.
However, in the DC network \eqref{sys:DC}, the interconnection structure is determined by the physical couplings of the circuit components, and the interconnected system \eqref{sys:DC} does not have an output consensus property~\eqref{ob:DC} in itself. Therefore, we need to design the controller \eqref{FB:DC} to enforce output consensus.
\red
\end{secrem}

\section{Output Consensus for Linear Systems}\label{OCL:sec}
In this section, to expose the main ideas of this paper, we focus on linear systems, described by
\begin{align}\label{linear}
\left\{\begin{array}{l}
\dot x = A x + B u + d\\
y = H x,
\end{array}\right.
\end{align}
where $x \in \bR^n$ and $u, y \in \bR^m$ denote the state, input, and output, respectively, 
and $d \in \bR^n$ denotes a constant disturbance. 

Suppose that $B$ is of full column rank, and that there exist $P, Q \succ 0$ such that for $d = 0$ and $V(x) := |x|_P^2$,
\begin{align}\label{passive:linear}
\dot V(x)  \le - |x|_Q^2 + y^\top u.
\end{align}
Note that this requirement can be relaxed into passivity and output detectability properties.
Under \eqref{passive:linear}, the controller \eqref{FB:DC} achieves output consensus, i.e., 
for some $\alpha \in \bR$,
\begin{align}\label{oc:linear}
\lim_{t \to \infty} y = \alpha \1_m.
\end{align}
In the following two subsections, we show this by using  two different approaches based on Krasovskii passivity and shifted passivity.

\subsection{Krasovskii Passivity Based Output Consensus}
First, we use Krasovskii passivity.
Thus, instead of $x$, we consider the dynamics of its time derivative  $\dot x$, and from (\ref{linear}) we obtain
\begin{align}\label{dlinear}
\left\{\begin{array}{l}
\ddot x = A \dot x + B \dot u \\
\dot y = H \dot x.
\end{array}\right.
\end{align}
Recall that $d$ is a constant, i.e., $\dot d = 0$.
Then, comparing \eqref{linear} with \eqref{dlinear}, one can observe from \eqref{passive:linear} that
\begin{align}\label{dpassive:linear}
\dot V( \dot x)  \le - |\dot x|_Q^2 + \dot y^\top \dot u.
\end{align}
This property is called strict Krasovskii passivity, which will be formally introduced in Definition \ref{def:Kpassive} in Section \ref{KP:sec}.

To study output consensus, we propose to use ${\bar V}_K (\dot x, y):= V(\dot x) + |y|_{\mathcal{L}}^2/2$ as an energy function.
Then, it follows from \eqref{FB:DC} and \eqref{dpassive:linear} that
\begin{align*}
\dot {\bar V}_K ( \dot x, y)  \le -  |\dot x|_Q^2 + \dot y^\top \dot u + \dot y^\top \mathcal{L} y = -  |\dot x|_Q^2.
\end{align*}
Since $Q \succ 0$, and $B$ is of full column rank, it holds from \eqref{FB:DC} and \eqref{dlinear} that
\begin{align*}
&\lim_{t \to \infty} \dot x (t)= 0
\; \implies \; 
\lim_{t \to \infty} \ddot x (t)= 0\\
&\qquad 
\implies \; 
\lim_{t \to \infty} \dot u (t)  = 0
\; \implies \; 
\lim_{t \to \infty} \mathcal{L} y (t)  = 0
\end{align*}
From  ${\rm ker}(\mathcal{L}) = {\rm span}\{ \1_m\}$, output consensus \eqref{oc:linear} is achieved.

\subsection{Shifted Passivity Based Output Consensus}
Next, we show output consensus from the viewpoint of shifted passivity by rewriting the controller \eqref{FB:DC} as
\begin{align}\label{FB2:DC}
\left\{\begin{array}{l}
\dot \xi =  E^\top y \\
u = - E \xi,
\end{array}\right.
\end{align}
where $\xi \in \bR^N$, and $E E^\top = \mathcal{L}$.
Computing $\dot u$, we obtain indeed \eqref{FB:DC}.

In the previous subsection, we have shown that $\lim_{t \to \infty} \dot x = 0$ and 
$\lim_{t \to \infty} \dot u = 0$. 
This implies that the closed-loop system consisting of \eqref{linear} and \eqref{FB2:DC} has 
an equilibrium point $(x^*, \xi^*) \in \bR^n \times \bR^m$.
Define $u^* := - E \xi^*$ and $y^* := H x^*$.
Then, the first equation of \eqref{FB2:DC} implies $E^\top y^* = 0$.
From \eqref{passive:linear}, one can verify that $V(x-x^*)$ satisfies 
\begin{align}\label{spassive:linear}
\dot V(x-x^*)  \le - |x-x^*|_Q^2 + (y-y^*)^\top (u - u^*).
\end{align}
This property is called strict shifted passivity, which will be formally introduced in Definition~\ref{def:spassive} in Section \ref{SP:sec}.

For ${\bar V}_s (x-x^*, \xi-\xi^*):= V(x-x^*) + |\xi-\xi^*|^2/2$,
it follows from \eqref{FB2:DC}, \eqref{spassive:linear}, $u^* = - E \xi^*$, and $E^\top y^*= 0$ that
\begin{align*}
\dot {\bar V}_s (x-x^*, \xi-\xi^*) 
& \le - |x-x^*|_Q^2 + (y-y^*)^\top (u - u^*) \\
&\,\quad + (y-y^*)^\top E (\xi - \xi^*) \\
&= - |x-x^*|_Q^2.
\end{align*}
Since $Q \succ 0$, and $B$ is of full column rank, it holds
from \eqref{linear}, \eqref{FB2:DC}, and $E E^\top = \mathcal{L}$ that
\begin{align*}
&\lim_{t \to \infty} x (t)= x^*
\; \implies \; 
\lim_{t \to \infty} \dot x (t)= 0\\
&\qquad 
\implies \; 
\lim_{t \to \infty} E \xi (t)= B^+ (A x^* + d)\\
&\qquad 
\implies \; 
\lim_{t \to \infty} E \dot \xi (t)= 0
\; \implies \; 
\lim_{t \to \infty} \mathcal{L} y (t)= 0.
\end{align*}
That is, output consensus \eqref{oc:linear} is achieved.

In the linear case, it is possible to obtain the consensus value~$\alpha$ and $x^*$ explicitly as
\begin{align}\label{alpha}
\alpha &= - \frac{ \1^\top (H A^{-1} B)^{-1} H A^{-1} d}{\1^\top (H A^{-1} B)^{-1} \1} \\
x^*
&= A^{-1} B (H A^{-1} B)^{-1} (\alpha \1 + H A^{-1} d)  - A^{-1} d,
\nonumber
\end{align}
and $\xi^*$ satisfies
\begin{align*}
E \xi^* = (H A^{-1} B)^{-1} (\alpha \1 + H A^{-1} d).
\end{align*}
One notices that $\alpha$ and $x^*$ are determined by $d$ only. 
In other words, the consensus value $\alpha$ is independent from
the controller dynamics \eqref{FB:DC} or \eqref{FB2:DC}.

In the above, we use the inverse of $H A^{-1} B$.
Its non-singularity  can be shown from \eqref{passive:linear}, i.e.,
\begin{align}\label{passive2:linear}
P A + A^\top P \preceq -  Q, \quad P B = H^\top
\end{align}
by contradiction.
If $H A^{-1} B$ is singular, there exists $v \in \bR^m \setminus \{0\}$ such that  $H A^{-1} B v = 0$.
It follows from \eqref{passive2:linear} that
\begin{align*}
v^\top B^\top A^{-\top}  Q A^{-1} B v
&\preceq 
- v^\top B^\top A^{-\top} (P A \\&\,\quad+ A^\top P  )A^{-1} B v\\
&= - 2 v^\top B^\top P A^{-1} B v\\
&= -2 v^\top H A^{-1} B v  = 0.
\end{align*}
This contradicts with $Q \succ 0$ or $B$ being of full column rank.

\begin{secrem}\label{rm:shifteq}
The consensus value $\alpha$ can be shifted by adding a constant input $\bar u$ to \eqref{FB2:DC} as follows:
\begin{align*}
\left\{\begin{array}{l}
\dot \xi =  E^\top y \\
u = \bar u - E \xi.
\end{array}\right.
\end{align*}
From \eqref{alpha}, the new consensus value $\alpha$ becomes
\begin{align*}
\alpha = - \frac{ \mathds{1}^\top (H A^{-1} B)^{-1} H A^{-1} (B \bar u + d)}{\mathds{1}^\top (H A^{-1} B)^{-1} \mathds{1}}
\end{align*}
That is, by a simple modification of the controller, one can control the consensus value $\alpha$ also.
\red
\end{secrem}

Now, we have shown that the controller \eqref{FB:DC} or equivalently \eqref{FB2:DC}
achieves output consensus based on Krasovskii passivity or shifted passivity.
As illustrated above, for Krasovskii passivity based output consensus, we do not assume that 
the closed-loop system has an equilibrium point beforehand.
This is an advantage in comparison with shifted passivity based output consensus, especially in the nonlinear case.
The shifted passivity based approach has also an own advantage that
the results can readily be generalized to time-varying cases, as will be explained later.

\section{Krasovskii Passivity Based Output Consensus}\label{KP:sec}
In this subsection, we generalize the Krasovskii passivity based approach to 
nonlinear systems, described by
\begin{align}\label{sys}
\left\{\begin{array}{l}
\dot x = f(x, u, d) \\
y = h(x,d),
\end{array}\right.
\end{align}
where $f:\bR^n \times \bR^m \times \bR^r \to \bR^n$ and $h:\bR^n \times \bR^r \to \bR^m$ are of class $C^1$, and
$\partial f(x,u,d)/\partial u$ is of full column rank at each $(x, u, d) \in \bR^n \times \bR^m \times \bR^r$.
We recall that $d \in \bR^r$ is a constant disturbance, i.e., $\dot d=0$.

As in the linear case, we use the extended system of \eqref{sys}:
\begin{align}\label{dsys}
\left\{\begin{array}{l}
\dot x = f(x, d, u) \\[2mm]
\displaystyle \frac{d\dot x}{dt} = \frac{\partial f(x,u, d)}{\partial x} \dot x 
+ \frac{\partial f(x,u, d)}{\partial u}  \dot u\\[2mm]
\displaystyle \dot y = \frac{\partial h(x, d)}{\partial x} \dot x.
\end{array}\right.
\end{align}
This can be understood as a system with the state $(x, \dot x, u)$, input $\dot u$, and output $\dot y$.
Focusing on the dynamics of $\dot x$,
we define Krasovskii passivity as a variant of the original definition \cite[Definition 2.8]{KKS:20}.

\begin{secdefn}\label{def:Kpassive}
Given $d \in \bR^r$, the system \eqref{sys} is said to be \emph{strictly Krasovskii passive} on $\mathcal{D} \subset \bR^n \times \bR^m$ if
for its extended system \eqref{dsys}, 
there exist $V_K: \mathcal{D} \times \bR^n \to \bR$ of class~$C^1$ and 
continuous $W_K: \mathcal{D} \times \bR^n \to \bR$ such that
\begin{align}
&V_K(x, u, \dot x) \ge 0, 
\label{cond1:Kpassive}\\
&W_K(x, u, \dot x) \ge 0 \mbox{ \; and \; } W_K(x, u, \dot x) = 0 \; \iff \; \dot x = 0
\label{W}\\
&\dot V_K(x, u, \dot x) \le -  W_K(x, u, \dot x) + \dot y^\top \dot u,
\label{cond2:Kpassive}
\end{align}
for all $(x, u) \in \mathcal{D}$ and $(\dot x, \dot u) \in \bR^n \times \bR^m$.
\red
\end{secdefn}

To deal with weighted output consensus, we generalize the controller \eqref{y:DC} as 
\begin{align}\label{FB:Kpassive}
\dot u = - M^\top \mathcal{L} M y,
\end{align}
where $\mathcal{L} \succeq 0$ is such that ${\rm ker}(\mathcal{L}) = {\rm span}\{ \1_m\}$, and 
$M\in \bR^{m \times m}$ is a matrix representing the weight associated with the output, which can be singular. For example, in DC microgrids it is generally desired that the total current demand is shared among the various nodes proportionally to the generation capacity of their corresponding energy sources (proportional current sharing).

Now, we present the first main result of this paper.
\begin{secthm}\label{thm:Kpassive}
Given $d \in \bR^r$, suppose that the closed-loop system consisting of a strictly Krasovskii  passive system \eqref{sys} on $\mathcal{D}$
and a controller \eqref{FB:Kpassive} is positively invariant on a compact set $\Omega \subset \mathcal{D}$.
Then, there exists some $\alpha: \bR \to \bR$ such that
\begin{align}\label{woc}
\lim_{t \to \infty} (M y(t) - \alpha (t) \mathds{1}_m)= 0
\end{align}
for each $(x(0), u(0)) \in \Omega$.
\end{secthm}

\begin{pf}
Define $\bar V_K(x, u, \dot x) := V_K(x, u, \dot x) + |M y|_\mathcal{L}^2/2$.
Then, it follows from \eqref{cond2:Kpassive} and \eqref{FB:Kpassive} that
\begin{align}\label{pf1:Kpassive}
\dot {\bar V}_K(x, u, \dot x)
&\le  - W_K (x, u, \dot x) + \dot y^\top \dot u + \dot y^\top M^\top \mathcal{L} M y \nonumber\\
&= - W_K (x, u, \dot x)
\end{align}
for all  $(x, u) \in \mathcal{D}$ and $\dot x \in \bR^n$.
Taking the time integration yields
\begin{align*}
&{\bar V}_K(x(t), u(t), \dot x(t)) 
+  \int_0^t W_K(x(\tau), u(\tau), \dot x(\tau)) d\tau\\
&\le {\bar V}_K(x(0), u(0), \dot x(0)).
\end{align*}
Since the closed-loop system is positively invariant on the compact set $\Omega$,
$\int_0^t W_K(x(\tau), u(\tau), \dot x (\tau)) d\tau$ exists for any $(x(0), u(0)) \in \Omega$ and 
$\dot x(0) \in \bR^n$.
Also, this is upper bounded and increasing with respect to $t \ge 0$, which implies that $\lim_{t \to \infty}\int_0^t W_K(x(\tau), u(\tau), \dot x (\tau)) d\tau$ exists and is finite.
Thus, it follows from \eqref{W}, Barbalat's lemma \cite[Lemma 8.2]{Khalil:96}, and 
the uniform continuity of $\dot x$ and $\ddot x$ on $\Omega$ that
\begin{align*}
&\lim_{t \to \infty}  W_K(x(\tau), u(\tau), \dot x (\tau)) = 0\\
&\qquad \iff \; 
\lim_{t \to \infty} \dot x (t) = 0
\; \implies \; 
\lim_{t \to \infty} \ddot x (t) = 0
\end{align*}
for any $(x(0), u(0)) \in \Omega$ and $\dot x(0) \in \bR^n$.
Therefore, it holds from \eqref{dsys}, \eqref{FB:Kpassive}, and $\partial f/\partial u$ being of full column rank that
\begin{align*}
\lim_{t \to \infty} \dot u (t) = 0
\; \iff \; 
\lim_{t \to \infty} M^\top \mathcal{L} M y(t) = 0
\end{align*}
for any $(x(0), u(0)) \in \Omega$.
That is, we have \eqref{woc}.
\qed
\end{pf}

\begin{secrem}\label{rem:con}
In Theorem \ref{thm:Kpassive}, $\alpha: \bR \to \bR$ is not necessarily to be a constant. 
To guarantee that the consensus value $\alpha$ is constant, we need an additional assumption such as the convergence of $x(t)$.
Such an assumption holds if $V_K(x, u, \dot x)=|\dot x |_P^2/2$, $P \succ 0$
and $W(x, u, \dot x)=|\dot x |_Q^2/2$, $Q \succ 0$.
In this case, \eqref{pf1:Kpassive} becomes
\begin{align*}
\frac{d (|\dot x (t)|_P^2 + |M y(t)|_\mathcal{L}^2)}{dt}
\le - | \dot x (t) |_Q^2.
\end{align*}
Taking the time integration yields
\begin{align*}
|\dot x (t)|_P^2 
&\le |\dot x (t)|_P^2 + |M y(t)|_\mathcal{L}^2\\
&\le |\dot x (0)|_P^2 + |M y(0)|_\mathcal{L}^2 -  \int_0^t  | \dot x (\tau ) |_Q^2 d\tau\\
&\le |\dot x (0)|_P^2 + |M y(0)|_\mathcal{L}^2 -  c  \int_0^t  | \dot x (\tau ) |_P^2 d\tau
\end{align*}
for some $c>0$. Then, the Gronwall-Bellman inequality, e.g., \cite[Lemma A.1]{Khalil:96} leads to
\begin{align*}
|\dot x (t)|_P^2 
&\le e^{- c t} (|\dot x (0)|_P^2 + |M y(0)|_\mathcal{L}^2 ).
\end{align*}
Therefore, the convergence speed of $\dot x(t)$ is exponential,
which implies that $x(t)$ converges to a constant. 
Accordingly, $y(t)$ converges to a constant, and the consensus value $\alpha$ is constant.
\red
\end{secrem}

\begin{secrem}\label{rem:rlx}
In Theorem \ref{thm:Kpassive}, we can relax strict Krasovskii passivity into Krasovskii passivity and 
a kind of output detectability.
Suppose that the system \eqref{sys} is Krasovskii passive, i.e., \eqref{cond1:Kpassive} and
\begin{align*}
\dot V_K(x, u, \dot x) \le \dot y^\top \dot u
\end{align*}
instead of \eqref{cond2:Kpassive}. Then, the controller $u = - K y + v$ with $K \succ 0$ achieves, for the closed-loop system,
\begin{align*}
\dot V_K(x, u, \dot x) \le - |\dot y|_K^2 + \dot y^\top \dot v.
\end{align*}
If the system \eqref{sys} has a kind of output detectability property:
\begin{align*}
\lim_{t \to \infty} \ddot u (t) = 0 
\mbox{ \; and \; } 
\lim_{t \to \infty} \dot y (t) = 0
\; \implies \; 
\lim_{t \to \infty} \dot x (t) = 0,
\end{align*}
then one can achieve weighted output consensus \eqref{woc} by $\dot v = - M^\top \mathcal{L} M y$.
This can be shown in a similar manner as the proof of Theorem \ref{thm:Kpassive}
based on the fact that $\dot y = 0$ implies $\ddot u = - K \ddot y - M^\top \mathcal{L} M \dot y = 0$.
\red
\end{secrem}

Theorem \ref{thm:Kpassive} illustrates that a simple controller \eqref{FB:Kpassive} achieves weighted output consensus.
However, its tuning parameter is only $\mathcal{L}$, and there is no enough freedom to improve control performances. 
Regarding this aspect, one can utilize the following dynamic extension:
\begin{align}\label{FB2:Kpassive}
\left\{\begin{array}{l}
\dot \rho = -\rho + y \\
\dot u = - M^\top \mathcal{L} M y + K (\rho - y),
\end{array}\right.
\end{align}
where $K \succ 0$ is a new tuning parameter.
Following similar steps as Theorem \ref{thm:Kpassive}, we can confirm that this new controller achieves weighted output consensus also.

\begin{seccor}\label{cor:Kpassive}
Given $d \in \bR^r$, suppose that the closed-loop system consisting of a strictly Krasovskii passive system \eqref{sys} on $\mathcal{D}$
and the controller \eqref{FB2:Kpassive} is positively invariant on a compact set $\Omega \subset \mathcal{D} \times \bR^m$.
Then, there exists some $\alpha: \bR \to \bR$ such that \eqref{woc} holds 
for each $(x(0), u(0), \rho (0)) \in \Omega$.
\end{seccor}
\begin{pf}
Define $\tilde V_K(x, u, \dot x, \rho) := V_K(x, u, \dot x) + |M y|_\mathcal{L}^2/2 + |\rho - y|_K^2/2$.
Then, it follows from \eqref{cond2:Kpassive} and \eqref{FB2:Kpassive} that
\begin{align*}
\dot {\tilde V}_K(x, u, \dot x, \rho)
&\le  - W_K (x, u, \dot x) + \dot y^\top \dot u + \dot y^\top M^\top \mathcal{L} M y \\
&\,\quad + (\dot \rho - \dot y)^\top K(\rho - y) \\
&= - W_K (x, u, \dot x) + \dot \rho^\top K(\rho - y) \\
&= - W_K (x, u, \dot x) - |\rho - y|_K^2
\end{align*}
for all  $(x, u) \in \mathcal{D}$, $(\dot x, \dot u) \in \bR^n \times \bR^m$, and $\rho \in \bR^m$.
The rest of the proof is similar to that of Theorem \ref{thm:Kpassive}.
\qed
\end{pf}

Theorem \ref{thm:Kpassive} and Corollary \ref{cor:Kpassive} can be generalized to impose partial weighted output consensus.
To achieve consensus among $y_j$, $j=i_1,\dots, i_{\bar m}$, $\bar m \le m$, 
one only has to implement the following controller:
\begin{align*}
\dot{\bar u} &= - \bar M^\top \bar{\mathcal{L}} \bar M \bar y, \\
\bar y 
&:= 
\begin{bmatrix} 
y_{i_1} & \cdots &y_{i_{\bar m}}
\end{bmatrix}^\top, \;
\bar u
:= 
\begin{bmatrix} 
u_{i_1} & \cdots &u_{i_{\bar m}}
\end{bmatrix}^\top,
\end{align*}
where $\bar{\mathcal{L}} \succeq 0$ is such that ${\rm ker}(\bar{\mathcal{L}}) = {\rm span}\{ \1_{\bar m}\}$, and 
$\bar M \in \bR^{\bar m \times \bar m}$ can be singular.
It is further possible to achieve weighted consensus among some of 
$y_j$, $j \neq i_1,\dots, i_{\bar m}$.

%\textcolor{blue}{It seems we can achieve kind of cluster consensus? E.g. y1-y3 converge to alpha1 and y4-y7 converge to alpha2. \\
%I guess in the overall communication topology (maybe unconnected), each $\bar L$ for cluster consensus should not be connected. \\
%As a different note, you may also make $\bar L$ or $L$ time-varying, but not sure how this complicates the analysis :).}

At the end of this section, we revisit the motivating example introduced in Section \ref{ME:sec}.
\begin{secex}
The DC network \eqref{sys:DC} satisfies 
\begin{align*}
\frac{d\dot x}{dt} =&~(\mathcal{J} - \mathcal{R}) \nabla^2 \mathcal{H} \dot x 
+
\begin{bmatrix}
0 \\ {\rm diag}\{ C_1 \frac{\dot q_1}{q_1^2}, \dots, C_n \frac{\dot q_n}{q_n^2}\} P_L^\star \\ 0
\end{bmatrix}\\
&+ g \dot u,
\end{align*}
where note that the Hessian matrix $\nabla^2 \mathcal{H}$ is constant.
Strict Krasovskii passivity can be shown by using $V_K (\dot x) =  |\dot x|_{\nabla^2 \mathcal{H}}^2 /2$.
Indeed, it follows that
\begin{align*}
\dot V_K ( \dot x)
=&~
\dot x^\top \nabla^2 \mathcal{H} (\mathcal{J} - \mathcal{R}) \nabla^2 \mathcal{H} \dot x  \\
&+
\dot x^\top \nabla^2 \mathcal{H}
\begin{bmatrix}
0 \\ {\rm diag}\{ C_1 \frac{\dot q_1}{q_1^2}, \dots, C_n \frac{\dot q_n}{q_n^2}\} P_L^\star \\ 0
\end{bmatrix} \\
&+ \dot x^\top \nabla^2 \mathcal{H} g \dot u\\
=&- W(\dot x, q) + \dot y^\top \dot u ,
\end{align*}
where $y$ is defined in \eqref{y:DC}, and 
\begin{align*}
W (\dot x, q)
:=&~\dot x^\top \nabla^2 \mathcal{H} \mathcal{R} \nabla^2\mathcal{H} \dot x \nonumber\\
&-
\dot x^\top \nabla^2 \mathcal{H}
\begin{bmatrix}
0 \\ {\rm diag}\{ C_1 \frac{\dot q_1}{q_1^2}, \dots, C_n \frac{\dot q_n}{q_n^2}\} P_L^\star \\ 0
\end{bmatrix}.
\end{align*}
Let $\mathcal{D}_q \subset \bR^\nu$ and $\Gamma \succ 0$ be such that
\begin{align*}
G_L^\star  - {\rm diag} \left(\frac{C_1^2 P_{L,1}^\star}{q_1^2}, \dots, \frac{C_n^2 P_{L,n}^\star}{q_n^2} \right) \succ \Gamma,
\quad \forall q \in \mathcal{D}_q,
\end{align*}
where $\Gamma$ represents a suitable lowerbound for the so-called equivalent conductance associated with the microgrid's loads.
Then, for $Q= {\rm diag} \{ L^{-1} R L^{-1} \allowbreak, \Gamma,\allowbreak L_t^{-1} R_t L_t^{-1}\}$, 
the DC network \eqref{sys:DC} is strictly Krasovskii passive on $\mathcal{D} = \bR^\nu \times \mathcal{D}_q \times \bR^\mu \times \bR^\nu$.

From Theorem \ref{thm:Kpassive},
the controller \eqref{FB:Kpassive} achieves weighted current sharing \eqref{woc} under 
the positive invariance assumption that is common in the literature on DC microgrids with constant power loads (see e.g. \cite{FCS:20} and the references therein). 
Furthermore, from Remark \ref{rem:con}, the consensus value $\alpha$ is constant. {We note that Krasovskii-like passivity has been already used for the design and analysis of voltage controllers for electric circuits and grids (see e.g. \cite{9029657,9093201}). However, to the best of our knowledge, Krasovskii passivity has never been exploited before for achieving current sharing and, more generally, output consensus.}
\red
\end{secex}

\section{Shifted Passivity Based Output Consensus}\label{SP:sec}
\subsection{Converging Disturbances}
In Section~\ref{OCL:sec}, it has been mentioned that
an advantage of shifted passivity based approach is the ease of dealing with a time-varying case.
To see this, we consider nonlinear time-varying systems, described by
\begin{align}\label{tvsys}
\left\{\begin{array}{l}
\dot x = f(t, x, d) + g(t, x, d) u \\
y = h(t, x, d),
\end{array}\right.
\end{align}
where $d: \bR \to \bR$ is a bounded continuous function of $t \in \bR$.
Also, $f:\bR \times \bR^n \times \bR^r \to \bR^n$, $g:\bR \times \bR^n  \times \bR^r \to \bR^{n \times m}$, and 
$h:\bR \times \bR^n  \times \bR^r  \to \bR^m$ are continuous in $(t, d)$ and locally Lipschitz in $x$
on $\bR \times \bR^n \times \bR^r$, and 
$g(t, x, d)$ is of full column rank at each $(t, x, d) \in \bR \times \bR^n \times \bR^r$. 

To deal with the time-varying case, 
we assume that the system~\eqref{tvsys} admits an equilibrium trajectory.
Namely, given $d(t)$, $t \in \bR$, there exists a class $C^1$ bounded trajectory $(x^*(t), u^*(t))$, $t \in \bR$, such that
\begin{align*}
\dot x^*(t) = f(t, x^*(t), d(t)) + g(t, x^*(t), d(t)) u^*(t), \quad
\forall t \in \bR.
\end{align*}

As a technical assumption, in this subsection, we assume a sort of convergence properties of $d$. 
This is relaxed in the next subsection.
\begin{secasm}\label{convd:asm}
Both $\lim_{t \to \infty} f(t, x^*(t), d(t))$ and $\lim_{t \to \infty} g(t, x^*(t), d(t))$ exist and are finite.
\red
\end{secasm}

To extend the concept of shifted passivity, e.g., \cite[Definition 2.14]{KKS:20} to the time-varying case,
we use the error dynamics $e := x - x^*$:
\begin{align}\label{tverr}
\left\{\begin{array}{l}
\dot e =  f(t, e + x^*, d) + g(t, e + x^*, d) u \\
\qquad - (f(t, x^*, d) + g(t, x^*, d) u^*) \\
y = h(t, e + x^*, d).
\end{array}\right.
\end{align}

\begin{secdefn}\label{def:spassive}
The system \eqref{tvsys} is said to be \emph{strictly shifted passive} 
along $(x^*(t), u^*(t))$ on $\mathcal{D} \subset \bR^n$ if 
for the error dynamics \eqref{tverr},
there exist $V_s: \bR \times \mathcal{D} \to \bR$ of class~$C^1$ and 
continuous positive definite $W_s:\mathcal{D} \to \bR$ such that
\begin{align}
&V_s(t, e) \ge 0, \\
&\dot V_s(t, e) \le -  W_s(e) + (y-y^*)^\top (u - u^*)
\label{cond2:spassive}
\end{align}
for all $(t, e) \in \bR \times \mathcal{D}$ and $u \in \bR^m$, 
where $y^* := h(t, x^*(t), d(t))$. 
When $W_s (\cdot ) = 0$, we simply say that the system is shifted passive.
\red
\end{secdefn}

As in the linear case, we use another representation of the controller \eqref{FB:Kpassive}:
\begin{align}\label{FB:spassive}
\left\{\begin{array}{l}
\dot \xi = E^\top M y \\
u = -  M^\top E \xi,
\end{array}\right.
\end{align}
where $\xi \in \bR^N$, and $E E^\top  = \mathcal{L}$. 
We show that this controller achieves weighted output consensus for a strictly shifted passive system.

\begin{secthm}\label{thm:spassive}
Given $d(t)$, $t \in \bR$, suppose that 
\begin{enumerate}
\renewcommand{\labelenumi}{\arabic{enumi})}

\item the closed-loop system consisting of
a system \eqref{tvsys} and a controller \eqref{FB:spassive} 
admits a class $C^1$ bounded trajectory $(x^*(t), \xi^*(t) )$, $t \ge 0$, such that 
$\dot \xi^*= E^\top M y^* = 0$, i.e., $\xi^*$ is constant;

\item for $(x^*(t), \xi^* )$ in item 1), the system \eqref{tvsys} satisfies Assumption \ref{convd:asm} and is strictly shifted passive along $(x^*(t), u^* )$ on $\mathcal{D}$, where $u^* := -  M^\top E \xi^*$;

\item when rewriting the system \eqref{tvsys} as the error dynamics \eqref{tverr}, 
the closed-loop system is positively invariant 
on some compact set $\Omega \subset \mathcal{D} \times \bR^N$ (for any initial time $t_0 \in \bR$), where the projection of $\Omega$ onto the $e$-space contains the origin.
\end{enumerate}
Then, there exists some $\alpha :\bR \to \bR$ such that \eqref{woc} holds 
for any $t_0 \in \bR$ and $(e(t_0), \xi(t_0)) \in \Omega$.
\end{secthm}

\begin{pf}
Define $\bar V_s (t, e, \xi) := V_s(t,e) + | \xi - \xi^*|^2/2$.
Then, it follows from \eqref{cond2:spassive}, \eqref{FB:spassive}, $\dot \xi^* = E^\top M y^*=0$ in item 1), and $u^* = - M^\top E \xi^*$ in item 2) that
\begin{align*}
\dot {\bar V}_s (t, e, \xi)
\le& -  W_s(e) + (y-y^*)^\top (u - u^*)  \\
&+ (y - y^*)^\top M^\top E (\xi - \xi^*) = -  W_s(e)
\end{align*}
for all $t \in \bR$ and $(e, \xi) \in \Omega$.
Taking the time integration yields
\begin{align*}
{\bar V}_s (t, e(t), \xi (t))
+ \int_{t_0}^t W_s(e (\tau) ) d\tau 
\le 
{\bar V}_s (t_0, e(t_0), \xi (t_0)).
\end{align*}
Since $\Omega$ is positively invariant with respect to $(e(t), \xi(t))$, and 
since from item 3), the projection of $\Omega$ onto the $e$-space contains the origin,
taking $t \to \infty$ and using Barbalat's lemma leads to
\begin{align*}
&\lim_{t \to \infty} W_s(e (t) )= 0 
\; \iff \; 
\lim_{t \to \infty} e (t) = 0\\
&\qquad \implies \; 
\lim_{t \to \infty} \dot e (t) = 0 
\end{align*}
for any $t_0 \in \bR$ and $(x(t_0), \xi (t_0)) \in \Omega$.

It follows from \eqref{tverr}, \eqref{FB:spassive}, Assumption~\ref{convd:asm}, 
$\dot \xi^*=0$ in item 1), the continuity of $f$ and $g$, $g$ being of full column rank, and $E E^\top = \mathcal{L}$ that
\begin{align*}
&\lim_{t \to \infty} ( f(t, e + x^*, d) - g(t, e + x^*, d) M^\top E \xi \\
&\qquad - (f(t, x^*, d) - g(t, x^*, d) M^\top E \xi^*)) \\
&= \lim_{t \to \infty} g(t, x^*, d) M^\top E (\xi - \xi^*) = 0 \\
&\implies \; 
\lim_{t \to \infty} M^\top E (\xi - \xi^*) = 0\\
&\implies \; 
\lim_{t \to \infty} M^\top E \dot \xi (t) 
= \lim_{t \to \infty}  M^\top \mathcal{L} M y (t)  = 0,
\end{align*}
which completes the proof.
\qed
\end{pf}

\begin{secrem}\label{rem:spssive}
The controller dynamics \eqref{FB:spassive} look similar to the edge dynamics imposed in  \cite{BZA:14, BD:15, MD:17}.
However, ours is more general in the sense of that $E$ is not necessarily to be an incidence matrix, and 
ours is weighted by $M$ to achieve weighted output consensus.
More importantly, we only require shifted or Krasovskii passivity of the system and 
can handle a wider class of systems than \cite{BZA:14, BD:15, MD:17}.
Namely, we reveal that imposing special edge dynamics is not an essential requirement to achieve output consensus when
one takes control design into account.
\red
\end{secrem}

\begin{secrem}
A similar remark as Remark \ref{rem:rlx} holds.
Namely, in Theorem~\ref{thm:spassive}, we can relax strict shifted passivity into shifted passivity and 
a kind of output detectability.
\red
\end{secrem}

\begin{secrem}\label{classK:rem}
The positive invariance assumption, i.e., item~3), can be removed from Theorem \ref{thm:spassive} if
there exist class $\cK$ functions $\alpha_1, \alpha_2$ such that
\begin{align*}
\alpha_1(|e|) \le V_s(t,e) \le \alpha_2 (|e|), 
\quad 
\forall t \in \bR, \; \forall e \in \mathcal{D}.
\end{align*}
Furthermore, global output consensus can be achieved if $\mathcal{D} = \bR^n$, and
$\alpha_1, \alpha_2$ are class $\cK_\infty$ functions.
\red
\end{secrem}

In the proof of Theorem~\ref{thm:spassive}, we show $\lim_{t \to \infty} e(t) = 0$, i.e., $\lim_{t \to \infty} ( x(t) - x^*(t)) = 0$.
This implies that the consensus value (more precisely, the valued-function):
\begin{align*}
\alpha (t) = \frac{\1_m^\top M h(t,x^*,d)}{\1_m^\top \1_m}
\end{align*}
does not depend on the initial state $(x(t_0), \xi(t_0))$ (or initial time $t_0 \in \bR$).
This further implies that if $x^*$ is constant, and $h$ is time-invariant, $\alpha$ is constant as well.

\subsection{More General Disturbances}
To deal with non-converging disturbances, we modify the controller \eqref{FB:spassive} as follows:
\begin{align}\label{FB2:spassive}
\left\{\begin{array}{l}
\dot \xi =  E^\top M y \\
u = - M^\top E (\xi + G E^\top M y),
\end{array}\right.
\end{align}
where $G \succ 0$ is a tuning parameter. 

Then, we have weighted output consensus without Assumption~\ref{convd:asm} for the convergence of the disturbance.

\begin{secthm}\label{thm2:spassive}
Given $d(t)$, $t \in \bR$, suppose that 
\begin{enumerate}
\renewcommand{\labelenumi}{\arabic{enumi})}

\item the closed-loop system consisting of a system \eqref{tvsys} and a controller \eqref{FB2:spassive} admits a class $C^1$ bounded trajectory $(x^*(t), \xi^*(t))$, $t \ge 0$ such that $\dot{\xi}^*= E^\top M y^* = 0$, i.e., $\xi^*$ is constant;

\item for $(x^*(t), \xi^* )$ in item 1), the system \eqref{tvsys} is 
shifted passive along $(x^*(t), u^*(t) )$ on $\mathcal{D}$, where $u^*(t) := - M^\top E (\xi^* + G E^\top M y^*(t))$;

\item item 3) of Theorem~\ref{thm:spassive} holds.
\end{enumerate}
Then, there exists some $\alpha :\bR \to \bR$ such that \eqref{woc} holds 
for any $(e(t_0), \xi(t_0)) \in \Omega$ and $t_0 \in \bR$.
\end{secthm}

\begin{pf}
Define $\bar V_s (t, e, \xi) :=  V_s(t,e) + | \xi - \xi^*|^2/2$.
Then, it follows from \eqref{cond2:spassive}, \eqref{FB2:spassive}, $\dot \xi^*= E^\top M y^*=0$ in item 1),
and $u^* = - M^\top E (\xi^* + G E^\top M y^*)$ in item 2) that
\begin{align*}
\dot {\bar V}_s (t, e, \xi) 
\le \;& (y-y^*)^\top (u - u^*)\\
&+ (y - y^*)^\top M^\top E (\xi - \xi^*) \\
=& - |E^\top M y|_G^2 - (y-y^*)^\top M^\top E (\xi - \xi^*) \\
&+ (y - y^*)^\top M^\top E (\xi - \xi^*) \\
=& - |E^\top M y|_G^2
\end{align*}
for all $t \in \bR$ and $(e, \xi) \in \Omega$.
In a similar manner as the proof of Theorem~\ref{thm:spassive}, we have
\begin{align*}
\lim_{t \to \infty} E^\top M y(t) =0 
\end{align*}
for any $(x(t_0), \xi (t_0)) \in \Omega$ and $t_0 \in \bR$.
\qed
\end{pf}

Theorem~\ref{thm2:spassive} can be generalized to non input-affine systems because we directly prove the convergence of $E^\top M y(t)$ without using the input-affine structure.

\begin{secrem}
A similar structure as \eqref{FB2:spassive} is found in \cite{BD:15, MD:17}, and
a similar remark as Remark \ref{rem:spssive} holds. 
Namely, our controller works for more general systems and for weighted output consensus. 
Moreover, we do not require that disturbance $d(t)$ is generated by an exosystem, nor the controller \eqref{FB2:spassive} satisfies the Sylvester-type equation for output regulation.
\red
\end{secrem}

\begin{secrem}
The positive invariance assumption can be removed from Theorem \ref{thm2:spassive} if
the conditions in Remark \ref{classK:rem} hold.
Furthermore, global output consensus can be achieved under
the same conditions in Remark \ref{classK:rem}.
\red
\end{secrem}

\begin{secrem}
As in Theorem \ref{thm:Kpassive} for Krasovskii passivity based output consensus, one can utilize the following dynamic extension of the controller~\eqref{FB2:spassive} to improve control performances:
\begin{align}\label{FB3:spassive}
\left\{\begin{array}{l}
\dot \rho =  - (\rho - y) \\
\dot \xi =  E^\top M y \\
u = - M^\top E (\xi + G E^\top M y) - K \rho,
\end{array}\right.
\end{align}
where recall that $K \succ 0$ is a new tuning parameter. 
One can confirm weighted output consensus when $y^*$ is constant, by utilizing a storage function
$\tilde V_s (t, e, \xi, \rho ) :=  V_s(t,e) + | \xi - \xi^*|^2/2 + | \rho - y^*|_K^2/2$.
A similar remark holds for Theorem \ref{thm:spassive} when $G = 0$.
\red
\end{secrem}

At the end of this subsection, 
we again revisit the motivating example.
\begin{secex}
Consider the system~\eqref{sys:DC} and controller~\eqref{FB2:spassive}.
From the convexity of $\mathcal{H}(x)$, it follows that
\begin{align}\label{H0}
\mathcal{H}_s (e, x^*) := \mathcal{H}(e + x^*) - \mathcal{H}(x^*) - \nabla^\top \mathcal{H}(x^*) e
\ge 0
\end{align}
for any $e, x^* \in \bR^{2\nu + \mu}$; 
this is a standard technique for shifting a storage function as found in \cite{JOG:07}.
Moreover, $\mathcal{H}$ in \eqref{sys:DC} satisfies $\mathcal{H}_s (e, x^*) = 0$ if and only if $e = 0$.
It follows from $x = e + x^*$, \eqref{H0}, and the structure of $\mathcal{H}$ in \eqref{sys:DC} that
\begin{align*}
\dot{\mathcal{H}}_s(e, x^*)
=&~(\nabla^\top \mathcal{H}(e + x^*)  - \nabla^\top \mathcal{H}(x^*)) (\dot e + \dot x^*) \\
&- e^\top \nabla^2 \mathcal{H} \dot x^*\\
=&~e^\top \nabla^2 \mathcal{H} \dot e\\
=&~e^\top \nabla^2 \mathcal{H} ( f(e + x^*) - f(x^*) + g (u-u^*) ) \\
=&- W_s(e, x^*) + (y - y^*)^\top (u-u^*) ,
\end{align*}
where  $y$ is defined in \eqref{y:DC}, and  
\begin{align*}
&W_s(e, x^*)\\
&:= -e^\top \nabla^2 \mathcal{H} ( f(e + x^*) - f(x^*))\\
&\; = e^\top \nabla^2 \mathcal{H} R \nabla^2 \mathcal{H} e\\
&\quad + e^\top \nabla^2 \mathcal{H}
\begin{bmatrix}
0 \\ ( {\rm diag} (C^{-1} q)^{-1} - {\rm diag} (C^{-1} q^*)^{-1}) P_L^\star \\ 0
\end{bmatrix}\hspace{-0.1cm}.
\end{align*}
Let $\mathcal{D}_q \subset \bR^\nu$ and $\Gamma \succ 0$ be such that
\begin{align*}
G_L^\star  - {\rm diag} \left(\frac{C_1^2 P_{L,1}^\star}{(e_1 + q_1^*) q_1^*}, \dots, \frac{C_\nu^2  P_{L,\nu}^\star}{(e_\nu + q_\nu^*) q_\nu^*} \right) \succ \Gamma,\\
\quad 
\forall t \ge t_0, \; \forall t_0 \in \bR, \; \forall e \in \mathcal{D}_q,
\end{align*}
where $e_i = q_i - q_i^*$, $i=1, \dots, \nu$.
Then, the DC network \eqref{sys:DC} is strictly shifted passive on $\mathcal{D} = \bR^\nu \times \mathcal{D}_q \times \bR^\mu$, where
$Q= {\rm diag} \{ L^{-1} R L^{-1}, \Gamma,\allowbreak L_t^{-1} R_t L_t^{-1}\}$.

From Theorem \ref{thm2:spassive}, 
the controller \eqref{FB2:spassive} achieves weighted current sharing under the positive invariance assumption.
When $q_i^*$, $i=1,\dots,n$ are constant, we can use the controller \eqref{FB:spassive}. 
In this case, the output consensus on $\mathcal{D}$ is guaranteed by the reasoning mentioned in Remark \ref{classK:rem}.
\red
\end{secex}

\section{Simulations}\label{sec:sim}
\begin{figure}[t]
\begin{center}
\begin{tikzpicture}[scale=0.76,transform shape,->,>=stealth',shorten >=1pt,auto,node distance=3cm,
                    semithick]
  \tikzstyle{every state}=[circle,thick,fill=white,draw=black,text=black]

  \node[state] (A)                    {Node 1};
  \node[state]         (B) [above right of=A] {Node 2};
  \node[state]         (D) [below right of=A] {Node 4};
  \node[state]         (C) [below right of=B] {Node 3};

  \path (A) edge   [below] node {~~~1} (B)
  		edge 	     node {4} (D)
           (B) edge      [below]        node {2~~~~} (C)
           (C) edge         [above left]     node {3} (D);

           \path[<->] (A) edge [bend left, dashed, blue]          node {} (B)
  		%edge	 [bend right, dashed, red]	     	node {} (D)
           (B) edge [bend left, dashed, blue]          	node {} (C)
           (C) edge [bend left, dashed, blue]          	node {} (D);
\end{tikzpicture}
\caption{Scheme of the considered microgrid composed of 4 nodes. The black solid arrows indicate the positive direction of the currents through the power lines. The dashed blue lines represent the communication network.}
\label{fig:microgrid_example}
\end{center}
\end{figure}
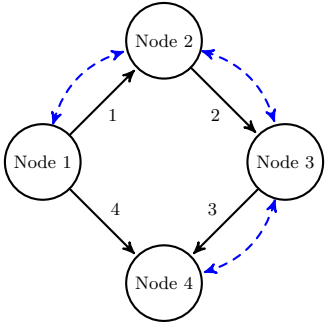

In this section, the proposed controllers are verified in simulation. We consider an islanded DC microgrid composed of 4 nodes in ring topology as shown in Figure~\ref{fig:microgrid_example}, where the dashed blue lines represent the communication network. As mentioned above, this system is both Krasovskii and shifted passive. The values of the parameters of
each node and line are mainly taken from \cite[Tables II, III]{CTD:18}, while those of the nominal loads are reported in Table~\ref{tab:parameters1}. Note that we consider also load variations, which are gathered into the disturbance $d$. For the sake of notational  simplicity, let $V_i:={q_i}/{C_i}$ and $I_i:={\varphi_i}/{L_i}$ denote respectively the voltage and the generated current associated with node $i=1,\dots,4$. The desired voltage value at each node is chosen equal to $V_i^*=$ \SI{380}{\volt} for all $i$. 
% The load components are as follows: $G_L^\star=\diag(0.1, 0.167, 0.25, 0.125)$ \SI{}{\siemens}, $I_L^\star=0$ \SI{}{\ampere}, and $P_L^\star=[0.25, 0.25, 0.25, 0.25]^\top$ \SI{}{\kilo\watt}. We consider $\delta(t) = 0$ in the interval $0\leq t < 10$ \SI{}{\second} and a step variation of the P-load of node 1 equal to $\Delta P_{L1}^\star =$ \SI{0.75}{\kilo\watt} at the time instant $t=$ \SI{5}{\second}. Then, we consider $\delta_{q1}(t) = - 0.1\frac{C_1}{q_1} \sin(t)$ \SI{}{\kilo\watt}, $\forall \, t\geq$ \SI{10}{\second}, which is equivalent to consider a \emph{time-varying} P-load in node 1. 
The controller parameters in \eqref{FB2:Kpassive} and \eqref{FB3:spassive} are chosen as $M=$ \num{100}$\,I_4$, $K =$ \num{0.2}$\,I_4$, $G = I_3$. Moreover, for the considered application, we select $\bar u$ in Remark~\ref{rm:shifteq} as $\bar u = C^{-1}q^* + RL^{-1}\varphi$. This simply allows us to shift the system equilibrium such that voltage average ($V_{\mathrm{av}}$) is equal to the voltage reference (see, e.g., \cite{TCC:19}). Four different scenarios are investigated in the following.\\

\begin{table}[t]
	\caption{Load Parameters}
	\centering
	{\begin{tabular}{lc | cccc}			
			Node								&	&1		&2	&3	&4\\			
			\hline
% 			$R_{i}$	&(\si{\ohm})	&\num{0.2}		&\num{0.3}		&\num{0.5}		&\num{0.1}\\
% 			$L_{i}$	&(\si{\milli\henry})	&\num{1.8}		&\num{2.0}		&\num{3.0}		&\num{2.2}\\
% 			$C_{i}$	&(\si{\milli\farad})	&\num{2.2}		&\num{1.9}		&\num{2.5}		&\num{1.7}\\
% 			$V_i^{*}$ &(\si{\volt})		&\num{380}		&\num{380}		&\num{380}		&\num{380}\\
			$P_{Li}^{\star}$ &(\si{\kilo\watt})		&\num{1}		&\num{2.5}		&\num{1.5}		&\num{5}\\
			$G_{Li}^{\star}$ &(\si{\siemens})	&\num{0.08}		&\num{0.04}		&\num{0.02}		&\num{0.08}\\
			$I_{Li}^\star$ &(\si{\ampere})	&\num{12.5}		&\num{7.5}		&\num{5.0}		&\num{15.0}\\
			$\Delta P_{Li}^\star$ &(\si{\kilo\watt})	&\num{4}		&\num{1}		&\num{1}		&\num{-4}\\
			%$\Pi_i$ &(\si{\kilo\watt})		&\num{25}		&\num{25}		&\num{25}		&\num{25}\\
	\end{tabular}}
	\label{tab:parameters1}
	\end{table}

\begin{figure}
\begin{center}
\includegraphics[width=\columnwidth]{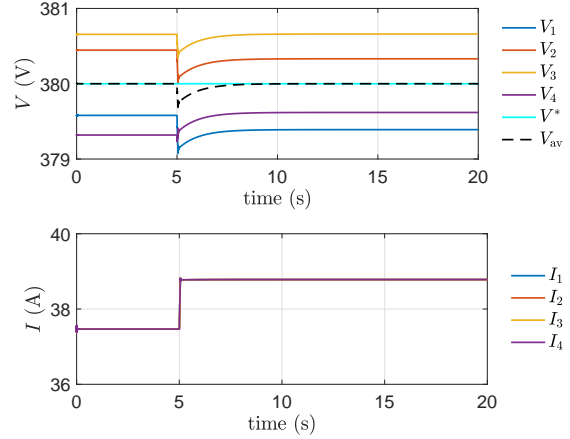}   
\caption{Scenario 1: current sharing with constant loads. {\bf(Top)} Time evolution of  the voltages and their average (dashed line) together with the corresponding reference (cyan line). {\bf(Bottom)} Time evolution of the currents.} 
\label{fig:0_0}
\end{center}
\end{figure}

{\bf Scenario 1.} In this scenario, we show current sharing with constant loads. Let the system initially be at the equilibrium. Then, at the time instant $t=$ \SI{5}{\second} a step variation equal to $\Delta P_L^\star$ (see Table \ref{tab:parameters1}) occurs in the P loads. From Fig.~\ref{fig:0_0}, we can observe that both voltages and currents converge to a constant equilibrium, where the voltage average converges to the voltage reference, and output consensus (i.e., current sharing) is achieved. Specifically, we can observe that since the loads are constant, then the consensus value is constant as well.\\ 

\begin{figure}
\begin{center}
\includegraphics[width=\columnwidth]{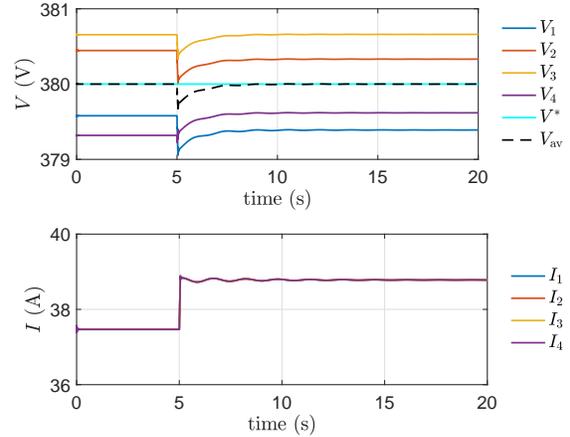}   
\caption{Scenario 2: current sharing with converging time-varying loads. {\bf(Top)} Time evolution of  the voltages and their average (dashed line) together with the corresponding reference (cyan line). {\bf(Bottom)} Time evolution of the currents.} 
\label{fig:1_0}
\end{center}
\end{figure}

{\bf Scenario 2.} In this scenario, we show current sharing with converging time-varying loads. Consider Scenario~1. At the time instant $t=$ \SI{5}{\second}, we add to the P load of node 3 a converging time-varying component equal to $0.1\mathrm{e}^{-0.25(t-5)}\sin(4t)$ \SI{}{\kilo\watt}.  From Fig.~\ref{fig:1_0}, we can observe that both voltages and currents converge to a constant equilibrium, where the voltage average converges to the voltage reference, and output consensus (i.e., current sharing) is achieved. Specifically, we can observe that since the loads converge to  constants, then the consensus value converges to a constant also.\\  

\begin{figure}
\begin{center}
\includegraphics[width=\columnwidth]{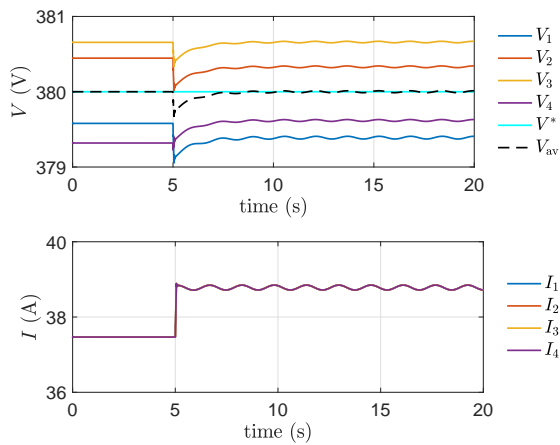}   
\caption{Scenario 3: current sharing with time-varying loads. {\bf(Top)} Time evolution of  the voltages and their average (dashed line) together with the corresponding reference (cyan line). {\bf(Bottom)} Time evolution of the currents.} 
\label{fig:2_0}
\end{center}
\end{figure}

{\bf Scenario 3.} In this scenario, we show current sharing with non-converging time-varying loads. Consider Scenario~1. At the time instant $t=$ \SI{5}{\second}, we add to the P load of node 3 a non-converging time-varying component equal to $0.1\sin(4t)$ \SI{}{\kilo\watt}. From Fig.~\ref{fig:2_0}, we can observe that both voltages and currents converge to an equilibrium trajectory, where the voltage average is stabilized around the voltage reference, and output consensus (i.e., current sharing) is achieved. Specifically, we can observe that since the loads are time-varying, then the consensus value depends on time.\\  

\begin{figure}
\begin{center}
\includegraphics[width=\columnwidth]{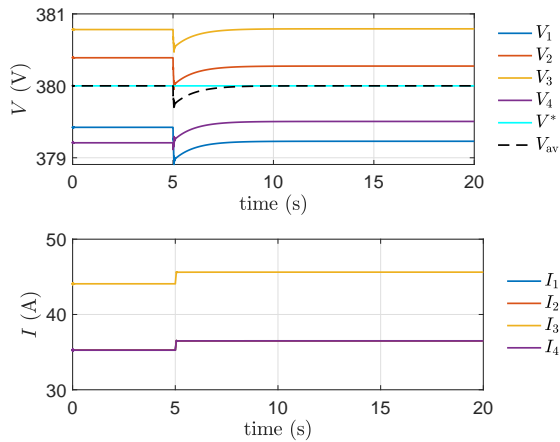}   
\caption{Scenario 4: weighted current sharing with constant loads. {\bf(Top)} Time evolution of  the voltages and their average (dashed line) together with the corresponding reference (cyan line). {\bf(Bottom)} Time evolution of the currents.} 
\label{fig:0_1}
\end{center}
\end{figure}

{\bf Scenario 4.} In this scenario, we show weighted current sharing with constant loads. Consider Scenario~1 with $M_{33}=$ \num{80}, which implies that node 3 is required to generate a current that is 25\% higher than the current generated by each of the other nodes. From Fig.~\ref{fig:0_1}, we can observe that both voltages and currents converge to a constant equilibrium, where the voltage average converges to the voltage reference, and weighted output consensus (i.e., proportional  current sharing) is achieved. Specifically, we can observe that $I_1=I_2=I_4$ and $I_3 = 1.25 I_1$.

\balance

\section{Conclusion}\label{Con:sec}
In this paper, we have studied an output consensus problem for nonlinear systems under external disturbances.
As the main contribution, we have proposed a simple distributed output feedback controller that 
achieves output consensus, based on Krasovskii or shifted passivity. 
An advantage of Krasovskii passivity based approach is that
we do not need to assume the existence of an equilibrium point for the closed-loop system.
An advantage of shifted passivity based approach is the ease of dealing with time-varying cases.
The utility of the proposed controller has been illustrated by an islanded DC power network
which is both Krasovskii and shifted passive.

\bibliographystyle{plain}        % Include this if you use bibtex 
\bibliography{output_consensus}           % and a bib file to produce the 
                                 % bibliography (preferred). The
                                 % correct style is generated by
                                 % Elsevier at the time of printing.

%\appendix
%\section{A summary of Latin grammar}    % Each appendix must have a short title.
%\section{Some Latin vocabulary}         % Sections and subsections are supported  
                                        % in the appendices.
\end{document}